\documentstyle[psfig]{mn}
\title{Galaxy Formation and Evolution.~II. \\
       Energy Balance,  Star Formation and Feed-Back}
\author[Fulvio Buonomo et al.]
        {Fulvio Buonomo$^{1}$, Giovanni Carraro$^{1}$,
        Cesare Chiosi$^{1}$, and Cesario Lia$^{2}$\\
       $^{1}$ Dipartimento di Astronomia, Universit\`a di Padova, Vicolo
dell'Osservatorio 5, I-35022 Padova, Italy \\
       $^{2}$ SISSA-ISAS, via Beirut 2-4, I-34013 Trieste, Italy \\ 
E-mail: {\tt buonomo,carraro,chiosi\char64pd.astro.it;liac\char64sissa.it}}

\date{\it Submitted: February 1998}
\pubyear{1999}
\begin{document}
\maketitle
\title{Star Formation and Feed-Back}

\begin{abstract}
In this paper we present a critical discussion of the algorithms
commonly used in N--body simulations of Galaxy Formation
to deal with the energy equation governing heating and cooling, 
to model star formation  and  the star formation rate, 
and to account for the  energy feed-back from stars. 
\par
First, we propose our technique for solving the energy equation in
presence of heating and cooling, which includes some difference 
with respect to the standard semi-implicit techniques.  
\par
Second, we examine the current criteria for  the onset of the
star formation activity. We suggest a new approach, in which star formation
is let
depend on the total mass density - baryonic (gas and stars) and dark matter 
- of the system  and  on  the metal-dependent   cooling  efficiency.
\par
Third, we check and discuss the separated effects of energy 
(and mass) feed-back from several sources - namely
supernovae, stellar winds from massive stars,  and UV flux from the same
objects.\\
All the simulations are performed in the framework of the formation
and evolution of a disk galaxy.\\
We  show that the inclusion of these physical phenomena
has a significant impact on the evolution of the galaxy models.
\par

\end{abstract}


\section{Introduction}
Numerical hydrodynamics and N--body simulations  are nowadays the
fundamental
tool to investigate galaxy formation and evolution.
The current status of numerical and semi-analytical 
modeling of  various galaxy properties 
has been recently  summarized by Frenk et al. (1997).

One of the poorly  understood processes  is the formation
of stars. Since the pioneering study by Katz (1992), 
the onset of star formation (SF) is usually empirically 
parameterized (see also Gerritsen 1997).
The key idea is that an element of fluid must satisfy some conditions
in order to be  eligible to SF and  turn part of its
gas content into stars according to a suitable star formation rate (SFR),  
which
is customarily a reminiscence of the Schmidt (1959) law. The reader is
referred to  the exhaustive 
discussion of this topic by Mihos \& Hernquist (1994) for more details.

The basic criteria to select the fluid  elements prone to SF  are
(i) the gas particle must be in a convergent flow, and (ii) the gas 
particle must be Jeans unstable.
The conditions are met if the velocity divergence is negative and if the
sound crossing time-scale is shorter than the dynamical time-scale.
 
Schmidt--like laws imply that $SFR \propto \rho^{n}$, where the power
$n$
ranges between 1 and 2. Conservation arguments suggest that the most probable
value of $n$ is $\frac{3}{2}$ (see Katz 1992). 
The  Schmidt law written for the volume density is 

\begin{equation}
\frac{d \rho_{\star}}{dt} = -\frac{d \rho_{g}}{dt} = \frac{c_{\star} 
\rho_{g}}{t_{g}} 
\end{equation}

\noindent
where $c_{\star}$ is the so-called  
dimensionless efficiency of star formation, 
and $t_{g}$ is the characteristic time
for the gas to flow. This is chosen to be the maximum between
the cooling time and the free-fall time $ t_{ff} = ( 4 \pi G 
\rho_{g})^{-\frac{1}{2}}$.

Over the past few years, starting from this general background a 
number of refinements have  been brought forth.
These mainly concern the criteria to decide whether a gas-particle is 
suitable to form stars, and  the way stars emerge from gas, i.e. how
gas clouds undergo fragmentation. \\

Navarro \& White (1993) and Katz, Weinberg \& 
Hernquist (1996) introduce an additional condition for the onset of SF, i.e. 
the so-called  over-density criterion, which secures that a 
collapsing cloud remains cool. The threshold density implies that 
the cooling time-scale is much shorter than the dynamical time-scale, 
and this always
occurs for temperatures greater than the cut-off temperature of the
cooling curve ($T = 10^{4} K$ for the typical mass resolution of this type
of problem). 
In addition to this, Katz, Weinberg \& Hernquist (1996) impose that 
the gas-particles can form stars only if 
a minimum physical density corresponding to 0.1 hydrogen
atoms per $\rm cm^{3}$ is met.

In Katz et al. (1996) and Navarro \& White (1993), 
the fragmentation of a gas cloud is usually modeled in the following
way. When in a gas-particle SF switches on, the gaseous mass decreases until
it reaches some minimum value, say 5\% of the original value. 
At this stage, the gas-particle 
is turned into a  collision-less star-particle, while the remaining gas 
is distributed among  the surrounding gas-particles. In the meantime 
the gas-particle is considered   as a dual entity, whose mass is partly
collision-less and partly collisional. 
This approach is a good compromise in terms of computational effort
because it is not necessary to add a collision-less particle every time
a new star is formed (Katz, Weinberg \& Hernquist 1996).
Then the  choice of the star forming regions is made by means of suitable 
probability arguments. Every time a gas-particle forms stars,
its mass is reduced by $1/3$ (free parameter).
  A different value is adopted by  Navarro \& White (1993),
who suppose  that at each star forming  event the gas mass is halved, and
that each gas-particle can split into  four star-particles  at most (this 
limit is set by  computational limitations).

A somewhat different scheme is followed by  Steinmetz (1995) who
 calculates the mass of the newly born
star-particle using equation (1) in the form:

\begin{equation}
m_{\star} = m_{g} (1 - exp(- \frac{c_{\star} \Delta t}{t_{ff}}) ) ,
\end{equation}

\noindent
where $\Delta t$ is the particle time step. 
The mass of the gas-particle is  accordingly reduced.
Although this latter approach appears  to be more physically
sounded, computational
problems can arise when many star-particles  are expected to form, so that
the number of star formation episodes must  be artificially limited.

Once  stars are present, 
they are expected to return to the interstellar medium (ISM) 
part of their mass (in form of chemically processed gas)
and energy via 
Supernovae (SN\ae) explosions, stellar winds,  and UV flux. These latter
contributions are significant only in the case of massive
stars. All this is known as the stellar energy Feed-Back (FB).

The amounts of mass, metals  and energy released by each star 
depend
on a star's mass, lifetime, and initial composition in a way
that can be easily taken into account.

The key problem here is to know how the  energy released by stars is given 
to the ISM because  the limited resolution of N--body simulations 
does not allow one to describe
the ISM as a multi-phase medium. Basically two different schemes exist:

(i) all the energy (from SN\ae, stellar winds, and   UV flux) is given to the
thermal budget of the fluid element (Katz 1992, Steinmetz \& M\"uller
1994); 

(ii) only a fraction of this 
energy is let kinetically affect the surrounding
fluid elements. This fraction ($f_{v}$ or $\epsilon_{kin}$) is a free
parameter
(Navarro \& White 1993, Mihos \& Hernquist 1994) chosen in such a way that
some macroscopic properties of the galaxy, 
like for instance the metallicity
distribution (Groom 1997) or the shape of the disk (Mihos \& Hernquist
1994), are reproduced.

The deposit of all the energy in the thermal budget 
of the fluid (first alternative) has little effect on a  galaxy' structure 
and evolution.
In fact, since  the thermal energy is given to a medium of high density and
short cooling time-scale in turn, it  is almost immediately
radiated away.
The difficulty cannot be cured by supposing that the thermal energy is
injected and shared over an e-folding time scale (Summers 1993). 

In all the above schemes the fraction of 
energy  from stellar FB which is not
transformed into kinetic energy
is added to the energy budget of the fluid element, and 
the fluid element is let cool.  
The  integration of the energy equation with respect of time 
is generally made according to a semi--implicit scheme
(see the discussion in Hernquist \& Katz 1989).

In this paper we present in detail our prescriptions for
computing energy balance, SF and FB.
The organization of the paper is  as follows.
Section~2 describes the numerical code and  the method used to integrate
the  energy equation, whereas Section~3 discusses in details the adopted
initial
conditions.
In section~4 we test our energy integration scheme following the formation
of a galactic disk.  
Section~5 presents tests a discussion on various SF criteria
, our new
approach to SF 
and the corresponding models.
Section~6 examines the effects of different sources of FB, while,
finally, some concluding remarks are presented in Section~7.

\section {The basic numerical tool}

\subsection{The Tree-SPH code}
All the simulations presented here  have been performed
using the Tree-SPH
code developed by Carraro (1996) and Lia (1996), and described  by 
Carraro
et al. (1998a), to whom the reader is referred  for all details.

The code, which can follow the evolution of a mix of Dark Matter (DM)
 and Baryons (gas and stars), 
 has been carefully checked against several classical
tests with  satisfactory  results, as reported in Carraro et al. (1998a).

In this code,  the properties of the gas component are 
followed by means of the Smoothed Particle
Hydrodynamics (SPH) technique (Lucy 1977, Gingold \& Monaghan 1997, Benz
1990), whereas the 
gravitational forces are computed
by means of the hierarchical tree algorithm of Barnes \& Hut (1986) using
a typical tolerance parameter $\theta=0.8$ and expanding tree nodes to quadrupole
order. We adopt a Plummer softening parameter.

In SPH each  particle represents a fluid element whose 
position, velocity, energy, density etc. are followed in  time and space.
The properties of the fluid are locally estimated by an interpolation which
involves the smoothing length $h_{i}$.
In our code  each particle possesses its own time and space 
variable smoothing
length $h_{i}$, and evolves with its own time-step.
This renders the code highly adaptive and flexible, and 
suited to speed-up the numerical calculations.

Radiative cooling is described by means of numerical tabulations as a
function of temperature and metallicity taken from Sutherland \& 
Dopita (1994).  This allows us to account for  the effects of variations in 
the metallicity among the  fluid elements and for each of these 
as a function of time and position. 

The chemical enrichment of the gas-particles caused by SF and 
stellar ejecta (Portinari et al 1998)
is described by means of the  closed-box model applied to
each gas-particle (cf. Carraro et al. 1998a,b for more details).

Star formation and Feed-back are discussed below.

Finally, all the calculations presented here 
have been carried out on  a DIGITAL ALPHA-2000 (330 Mhz of Clock, 512
Mbyte of RAM) workstation hosted by 
the Padua  Observatory \& Astronomy Department.

\subsection{On the integration of the energy equation}

The usual form of the energy equation  in SPH formalism is

\[
\frac{du_i}{dt}=\sum_{j=1}^N m_j \left
(\frac{\sqrt{P_{i}P_{j}}}{\rho_{i}\rho_{j}}
 + \frac{1}{2} \Pi_{ij}\right ){\bf v}_{ij} \times
\]

\begin{equation}
\frac{1}{2}
\left ( {\vec \nabla}_{i} W(r_{ij},h_i)+ \vec \nabla_i
W(r_{ij},h_j)\right ) + \frac {\Gamma - \Lambda_C}{\rho},
\label{eq_energy}
\end{equation}

\noindent
(Benz 1990; Hernquist \& Katz 1989). The first term
represents the heating or cooling  rate of mechanical nature, whereas
the second term  $\Gamma$ is  the total heating rate
from all sources apart from the mechanical ones, and the third term
 $\Lambda_C / \rho $ is the total
cooling rate by many physical agents (see Carraro et al. 1998a for
details).

In absence of explicit sources or sinks of energy the energy
equation
is adequately integrated using an explicit scheme and the Courant
condition for time-stepping (Hernquist \& Katz 1989).

The situation is much more complicated when considering cooling.
In fact, in real situations the cooling time-scale 
becomes much shorter 
than any other relevant  time-scale (Katz \& Gunn 1991),
and the time-step becomes considerably shorter than the Courant time-step. 
even  using the fastest  computers at disposal.
This fact makes it  impossible to integrate the complete system 
of equations (cf. Carraro et al. 1998a) adopting  as time-step the cooling
time-scale. 

To cope with this difficulty,
Katz \& Gunn (1991) damp the cooling rate to avoid too short
time-steps allowing gas particles to loose only half of their thermal energy
per time-step.
 
Hernquist \& Katz (1989) and Dav\'e et al. 1997 solve semi-implicitly eq.
(\ref{eq_energy}) using the trapezoidal rule,

\begin{equation}
u_{i}^{n+1} = u_{i}^{n} + \frac{1}{2} (dt_{i}^{n} \times e_{i}^{n} +
dt_{i}^{n+1} \times e_{i}^{n+1} ) ,  
\label{trapez}
\end{equation}

\noindent
where $e_{i} = du_{i}/dt$.
The leap-frog scheme is used to update thermal energy, and the energy
equation,
which is nonlinear for $u_{j}$, is solved iteratively both at the
predictor and at the corrector phase.
The technique adopted is a
hybrid scheme which is a combination of the  bi-section and Newton-Raphson
methods (Press et al 1989). The only assumption is that at the predictor
stage, when the predicted $\tilde u_{i}^{n+1}$  is searched for, the terms
$u_{i}^{n+1}$ are equal to $u_{i}^{n-1}$.\\
 
Our scheme to update energy is conceptually the same, but differs in 
the predictor stage and in the iteration scheme adopted to solve
equation \ref{trapez}.

In brief, at the first time-step the quantity  $u^{n-1/2}$ is calculated
and for all subsequent time steps, the leap-frog technique, 
as in Steinmetz \& M\"uller (1993), is used: 
\vskip 0.2cm

(i)  We start with $u^{n}$ at $t^{n}$;\\

(ii) compute $\tilde u^{n}$ as 
 
\[
~~~~~~~~~~\tilde u^{n} = u^{n-1/2} + \frac{1}{2} t^{n} \times e^{n}.
\]

\noindent
This predicted energy, together with the predicted velocity is used to
evaluated the viscous and adiabatic contribution to $e_{i}^{n+1}$.
In other words the predictor phase is calculated explicitly
because all the necessary quantities are available from the previous time
step $t^{n}$.\\

(iii) finally, derive $u^{n+1}$ solving the equation \ref{trapez}
iteratively (corrector phase) for both the predicted and old adiabatic and
viscous terms; \\

\noindent
In the corrector stage the integration of the equation \ref{trapez} 
is performed using the Brent method (Press et al 1989) instead of the
Newton-Raphson, the accuracy being fixed to a part in $10^{-5}$.
The Brent method has been adopted because it is better suited as
root--finder
for functions in tabular form (Press et al. 1989).

\section{Initial conditions}
In the following we are going to check our treatment of the
energy equation and our implementation of SF and FB simulating the
formation of a disk galaxy. Since our code is not a cosmological code
we shall consider an {\it ad hoc} initial configuration for a
protogalaxy. This approach is justified by the fact they we are going to 
focus on processes occurring at scales much lower than 
the cosmological ones.\\

\subsection{Theoretical overview}
According to modern gravitational instability theory (White \& Rees 1978),
protogalaxies comprise a mixture of dissipation-less dark matter (DM), whose
nature is not clear insofar, and dissipational baryonic material,
roughly  in the mass ratio 1:10.
After a violent relaxation process, which follows the separation from the
Hubble expansion, a DM halo becomes isothermal, and acquires baryonic
material which heats up at the halo virial temperature. 
Gas then cools and collapses, and through fragmentation turns into the
stars we see today in disks and spheroidal systems.\\

In this scenario disks form as a consequence of the angular momentum that
DM halos acquire due to the tidal torque felt by surrounding halos, while
spheroidal systems are thought to be produced by merging of disks.\\

Instead of selecting halos from cosmological N--body simulations (see for 
instance Weil et al 1998), in our code we set up a protogalaxy as an
isolated rotating DM halo with baryonic material inside,
proceeding as follows.

\begin{figure}
\centerline{\psfig{file=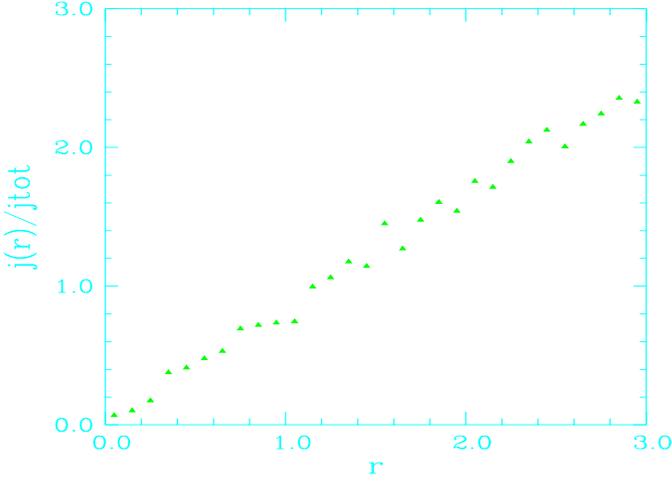,height=10cm,width=10cm}}
\caption{The initial specific angular momentum $j$ as a function of the
radial distance $r$ from the center of a $1/r$ dark matter halo.}
\end{figure}

\subsection{Initial configuration}
We consider a triaxial DM halo whose density profile is

\begin{equation}
\rho(r) \propto \frac{1}{r} .
\end{equation}

\noindent
Although rather arbitrary,
this choice seems to be quite reasonable. Indeed DM halos emerging
from cosmological N--body simulations are not King or isothermal spheres,
but show, independently from cosmological models, initial fluctuations
spectra and total mass, an {\it universal} profile (Navarro et al 1996; 
Huss et al 1998). 
This profile is not a power law, but has a slope $\alpha = dln\rho / d ln 
r$ with $\alpha = -1$ close to the halo center, and $\alpha = -3$ at
larger radii. Thus in the inner part the adopted profile matches the
{\it universal} one. Moreover this profile describes a situation which is
reminiscent of a collapse within an expanding universe, being the local
free fall time a function of the radius (see the discussion in Curir et 
al 1993; Aguilar $\&$ Merritt 1990 ).
Moreover triaxiality is quite natural for galactic DM halos 
(Becquaert \& Combes 1997; Olling \& Merrifield 1998).\\

DM particles are distributed by means of an {\it
acceptance-rejection} criterium, following step by step the procedure
developed by Curir et al (1993).
Specifically, we consider a triaxial ellipsoid whose radial scale $r$
is given by 

\begin{equation}
r = (x^{2} + \frac{y^{2}}{(b/a)^{2}} + \frac{z^{2}}{(c/a)^{2}})
\end{equation}

\noindent
where $a > b > c$ are the axial ratios (see Table~1).

\begin{table}
\tabcolsep 0.6truecm 
\caption{Initial conditions of the simulated Dark Matter halo:
N=10,000; a:b:c = 3.0: 2.25: 1.5.}
\begin{tabular}{cccc} \hline\hline
\multicolumn{1}{c}{$\lambda$} &
\multicolumn{1}{c}{$\beta$} &
\multicolumn{1}{c}{$\delta_{1}$} &
\multicolumn{1}{c}{$\delta_{2}$}\\ 
\hline
0.09& 0.20  & 0.05 & 0.025 \\
\hline
\end{tabular}
\end{table}

\noindent
The velocity field has been chosen to produce an angular momentum which
depends linearly on the distance, $j(r) \propto r$ (Barnes \& Efstathiou 1987).
It has been built up sampling the moduli $V$ of the
particles velocity by using a Maxwellian distribution and imposing that
the virial ratio $\beta = 2 \times T / |W|$ is equal to some
fixed value ranging between 0.05 and 0.20. Here $T$ is the kinetic energy,
while $W$ is the potential energy.
The final cartesian components of the particle velocity can be derived from
the following formulas

\begin{equation}
V_{x} = - V \times sin(\theta + \alpha)
\end{equation}

\begin{equation}
V_{y} =  V \times cos(\theta + \alpha)
\end{equation}

\begin{equation}
V_{z} = \xi
\end{equation}

\noindent
where $\theta$ is the angle between the position vector of a particle and the $x$ coordinate
axis. $\alpha$ is an angle varying between $- \delta_{1} \pi$ and $+ \delta_{1} \pi$,
whereas $\xi$ is a random parameter ranging from $- \delta_{2} V$ and $+ \delta_{2} V$.
$\delta_{1}$ and $\delta_{2}$ are chosen so that the kinetic energy $T$ changes by less than 
$1\%$ of the value giving the initial virial ratio $\beta$.\\
Table~1 summarizes the adopted initial values for the set of parameters introduced above.\\
Due to the assigned velocity field, the halo acquires an amount of angular momentum, which
is conventionally described by means of the dimensionless spin parameter $\lambda$:

\[
\lambda = \frac{J |E|^{1/2}}{G M^{5/2}}
\]

\noindent
Here $G$ is the gravitational constant, $J$ the system angular momentum, $M$
the total mass and $E$ the total system energy. In our case the $\lambda$ parameter
has chosen to be $0.09$, significantly greater than the mean values of cosmological
halos (Steinmetz \& Bartelmann 1995). This choice is motivated by the comparison
we are going to make with similar initial conditions (Navarro \& White 1993;
Thacker et al 1998; Raiteri et al 1996).\\
 
The softening parameter $\epsilon$ is computed as follows.
After plotting the inter-particles separation as a function of the distance
to the model center, we compute $\epsilon$ as the mean inter-particles
separation at the
center of the sphere, taking care to have at least one hundred particles
inside the softening radius (Romeo A. G., Pearce F. R., private
communications). We
consider a Plummer softening parameter, equal for both DM and gas
particles (see below),
and keep it constant along the simulation. For this particular choice
of the initial configuration the softening parameter $\epsilon$ turns out
to be $3.6 ~kpc$.\\

Total energy and angular momentum are conserved within $1\%$ and $0.1\%$
level, respectively.\\

We consider a dark matter halo with mass $10^{12} M_{\odot}$ and radius ($a$) 
$120 ~kpc$ in order to simulate a galaxy with a size similar to the Milky Way.
Axial ratios are $a = 3.00$, $b = 2.25$, $c = 1.50$.

\begin{figure*}
\centerline{\psfig{file=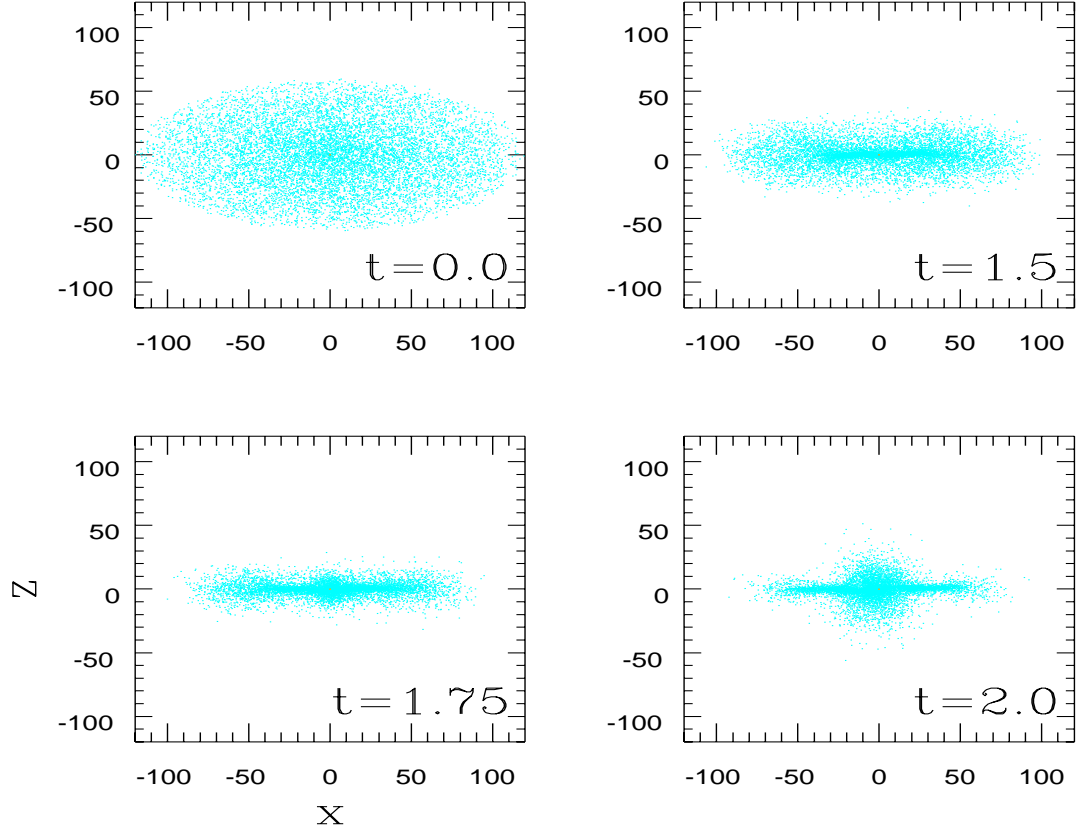,height=16cm,width=16cm}}
\caption{The formation process of a galactic disk 
in the $x-z$ plane. 
In the bottom--right corner of any snapshot time is reported in billion years.}
\end{figure*}

\begin{figure*}
\centerline{\psfig{file=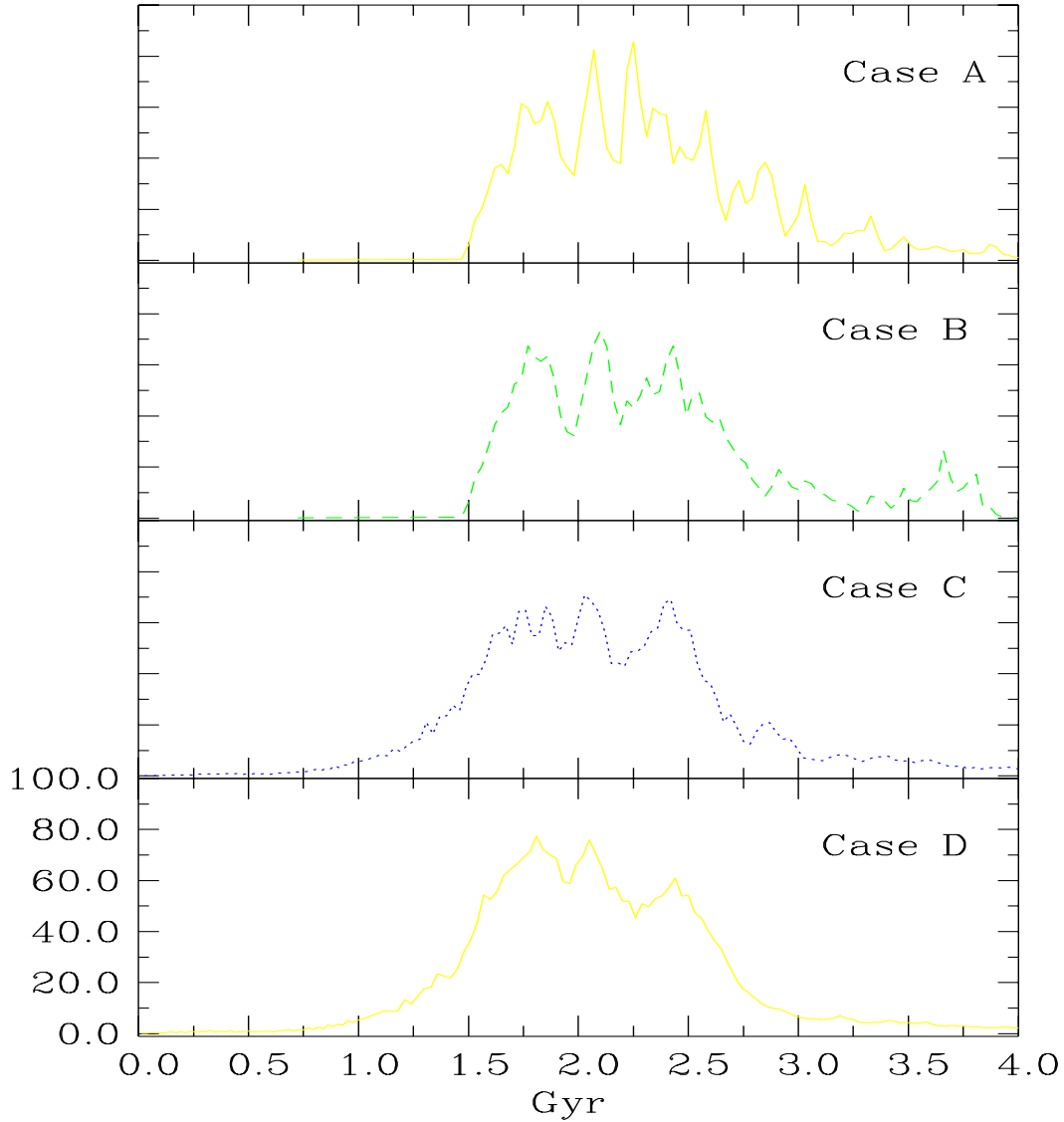,height=16cm,width=16cm}}
\caption{The Star Formation history ( in solar masses per year)
for four models in which star formation
criteria are defined as in Section.~5. See text for details}
\end{figure*}

\section{The formation of a gaseous disk} 
The formation of a galactic disk is
simulated distributing gas inside the halo described above and switching cooling on.
To mimic the in-fall of gas inside the potential well of the halo, 
we distribute gas particles (10,000 in number) on the top of DM particles.
The baryonic fraction adopted is $f_{b}~=0.1~$, and gas particles
are Plummer--softened in the same way as the DM particles.\\
  
Under cooling and the velocity field of the halo,  the gas is expected to settle down in a
rotating thin structure.
For this purpose we need to specify for the baryonic component a
temperature and a metal content.
We assume that the gas has a
temperature of $\approx 1.0
\times 10^{4}~^{o}K$ (Navarro \& White 1993), and an almost primordial metal 
content amounting  to $Z \approx 10^{-4}$, which translates into $[Fe/H]~\approx~-3$
(Bertelli et al 1994).\\

This simulation is meant to verify that our energy integration scheme
works properly when cooling is switched on. For this reason SF, FB and
chemical enrichment are turned off. \\

The formation of the disk is shown in Fig.~3.
Left panels follows from the top to the bottom the evolution of the gas component
in the $x-z$ plane. Time is in Gyr.\\
Starting from an ellipsoidal initial configuration
(t = 0.0), baryons  due to cooling and angular momentum settle down in a thin
rotating structure (t = 1.75), eventually developing a bulge like structure in the model
center (t = 2.5). At this time the gaseous disk is about $5~kpc$ high and 
$40~kpc$ wide.\\
Comparing our results with similar simulation by Raiteri et al (1996) we see that in our
case the disk forms more slowly. This is due to the lower efficiency 
of our cooling with respect to
the more widely adopted Katz \& Gunn (1991) cooling.

\section{Star formation}
As already recalled, SF in N--body simulations  is let occur
if
suitable criteria  meant  to simulate 
the real behaviour of the ISM are satisfied.

Current prescriptions for SF stand  on three  time-scales, i.e. 
the crossing, cooling,  and free-fall
time-scales, which ultimately depend (although in a different fashion) 
on the density of the fluid. According to the most popular prescription (Katz 1992;
Navarro \& White 1993),
SF is let occur if

\begin{description}
\item $\bullet$ $\nabla \cdot {\vec v_{i}}  < 0$
\item $\bullet$ $t_{sound} > t_{ff}$
\item $\bullet$ $t_{cooling} << t_{ff}$
\end{description}

\noindent
Aim of this section is to  closely  scrutiny 
the effects of three conditions on 
the model results. We start with some general considerations.  

Most likely, the divergence criterion
is not  strictly required. Firstly the typical mass resolution ($10^{6} - 10^{7} M_{\odot}$) 
of this kind of N-body simulation is clearly not sufficient to 
follow the kinematical behaviour of the gas at the molecular clouds scale. 
Then
in principle a fluid
element can form stars without being in a convergent flow (let us think about
shock induced or stochastic SF ).
For instance Mihos \& Hernquist (1994) adopt in the special case of an already
formed spiral galaxy a different dynamical criterium, based upon
the Toomre instability criterium.

The free-fall time scale is usually computed as 
\begin{equation}
t_{ff} = (4 \pi G \rho)^{-1/2}
\end{equation}
 
\noindent
where $\rho$ is the gas density of the fluid element.
We have replaced this condition with the more general one

\begin{equation}
t_{ff} = (4 \pi G \rho_{tot})^{-1/2}
\end{equation}

\noindent
where $\rho_{tot}$ = $\rho_{gas} + \rho_{DM}$. In principle,  one should
consider also $\rho_{star}$.
In fact the dynamical behaviour of the fluid element under
consideration depends on all the mass inside the volume with
radius $2 \times h$, where $h$ is the smoothing length.
Therefore at the typical resolution of these
simulations (about 2.5 kpc) the presence of DM accelerates
the collapse of a fluid element by lowering the free-fall time-scale.
DM density is evaluated in the same manner as gas density, defining
a smoothing length for any DM particle, 
although this scheme results
computationally expensive.\\
The importance of DM on SF has been recently 
studied from an analytical point of view by Caimmi \& Secco (1997).

Finally, we remind the reader  that  the condition on the cooling
time-scale 
is customarily replaced by a condition on the density, i.e. 
a constant density threshold (Navarro \& White 1993).
This  stems from suitable arguments about the 
cooling time-scale based on the notion that the gas metallicity has no 
effect. However, as the cooling rate does increase with  the metal content
of the ISM with consequent increase of the threshold density,
the more general condition on the cooling time-scale ought to be preferred.

\begin{table}
\tabcolsep 0.5truecm 
\caption{Comparison of SF criteria.}
\begin{tabular}{lccc} \hline
\multicolumn{1}{c}{Case} &
\multicolumn{1}{c}{SF peak  } &
\multicolumn{1}{c}{Peak time} &
\multicolumn{1}{c}{Stars formed} \\
\hline
& $M_{\odot} /yr$& $Gyr$& Number\\
\hline
{\bf A} &85 & 2.25 &3381 \\
{\bf B} &73 & 2.10 &3771 \\
{\bf C} &71 & 2.03 &6580 \\
{\bf D} &77 & 1.81 &8543 \\
\hline
\end{tabular}
\end{table}

\subsection{Testing the SF recipes}

To this aim we perform the same simulation as in Section~4, say the formation of a disk
galaxy, but letting
stars form according to different set of criteria.
We follow the model till the end to the major star formation episode.\\
The simulations have been made using  10,000 DM and 10,000 baryonic particles,
neglecting in this particular set of models 
any source of FB.
The following four cases are examined.

\vskip 0.3cm
\begin{center}
\begin{tabular}{|l c|c|c|}
\hline
 Case {\bf A}  &$\nabla \cdot {\vec v_{i}}  < 0$  &$t_{sound} > t_{ff}$  & $\rho > \rho_{crit}$ \\
 Case {\bf B}  &                                  &$t_{sound} > t_{ff}$  & $\rho > \rho_{crit}$ \\
 Case {\bf C}  &$\nabla \cdot {\vec v_{i}}  < 0$  &$t_{sound} > t_{ff}$  & $t_{cooling} << t_{ff}$\\
 Case {\bf D}  &                                  &$t_{sound} > t_{ff}$  & $t_{cooling} << t_{ff}$\\
\hline
\end{tabular}
\end{center}
\vskip 0.3cm

\noindent
In our models,  fragmentation of gas-particles undergoing SF is let occur
and new star-particles are created
when the gas content of the fluid element has fallen 
below $20\%$ of the initial value, being the remaininig
gas spread out over the surrounding particles.\\
This limits the number of star particles
produced.
Moreover the gas particle is allowed to
experiment  up to 
ten SF episodes. Afterwards it cannot make stars anymore.\\ 
In all simulations the efficiency parameter 
$c_{\star}$ is fixed to 0.1.
Finally, when more than two star-particles form inside a sphere with radius 
equal to the softening parameter $\epsilon = 3.6~kpc$,
they are merged together to form a single object, in order to keep
the total  number of particles small. 

\par
The results for the SFR as a function of time are shown in Fig.~3, whereas a few
relevant quantities of the models are given in Table~2.

\begin{figure}
\centerline{\psfig{file=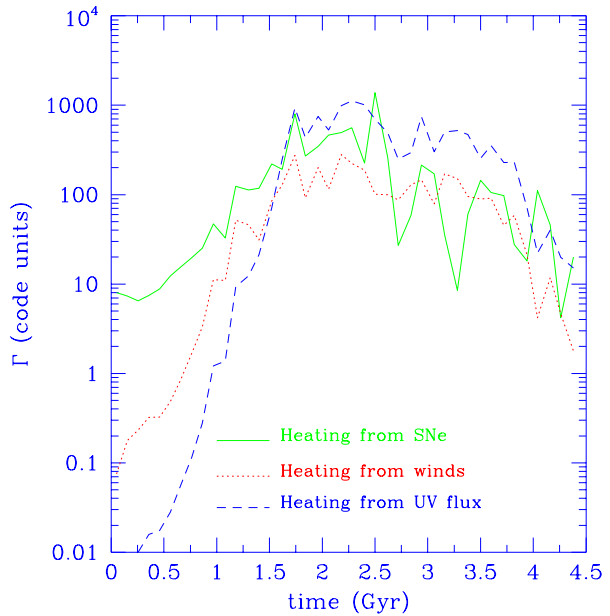,height=9cm,width=9cm}}
\caption{The heating rates  from the three sources of FB as indicated. See the 
text for more details. }
\end{figure}

\noindent
Comparing either case  {\bf A} to case  {\bf B} or case  {\bf C} to case 
{\bf D}, we notice that the criterion on the velocity divergence 
has in practice no effect, as argued above.\\ 
In contrast, there is a significant difference passing from
the cases  {\bf A} and  {\bf B} based on the over-density condition to 
the cases  {\bf C} and  {\bf D} based on the cooling time-scale.
In fact the models of the first group start sensibly later to form stars
and after the burst, which occurs somewhat later, SF maintains higher
that in the second group models, and show some secondary peaks.
Dropping the divergence criterium significantly increases the number of 
spawned stars (model {\bf C} and  {\bf D} ).\\
In these models after the peak SF proceeds with a rate around $4-6 ~M_{\odot}/yr$,
quite usual for disk galaxies.
Putting together all these differences and the discussion above we
are led to prefer the combination of criteria of the model {\bf D},
which we are going to use in all the other simulations
presented in this paper.\\
All simulations exhibit the SF peak at about $2~Gyr$, much later than, for instance,
the Raiteri et al (1996) simulation. This is clearly due to the different cooling
functions adopted. Katz $\&$ Gunn (1991) analytical formulas provide  a much more efficient 
cooling (see below and Carraro et al 1998a).
 
\begin{figure*}
\centerline{\psfig{file=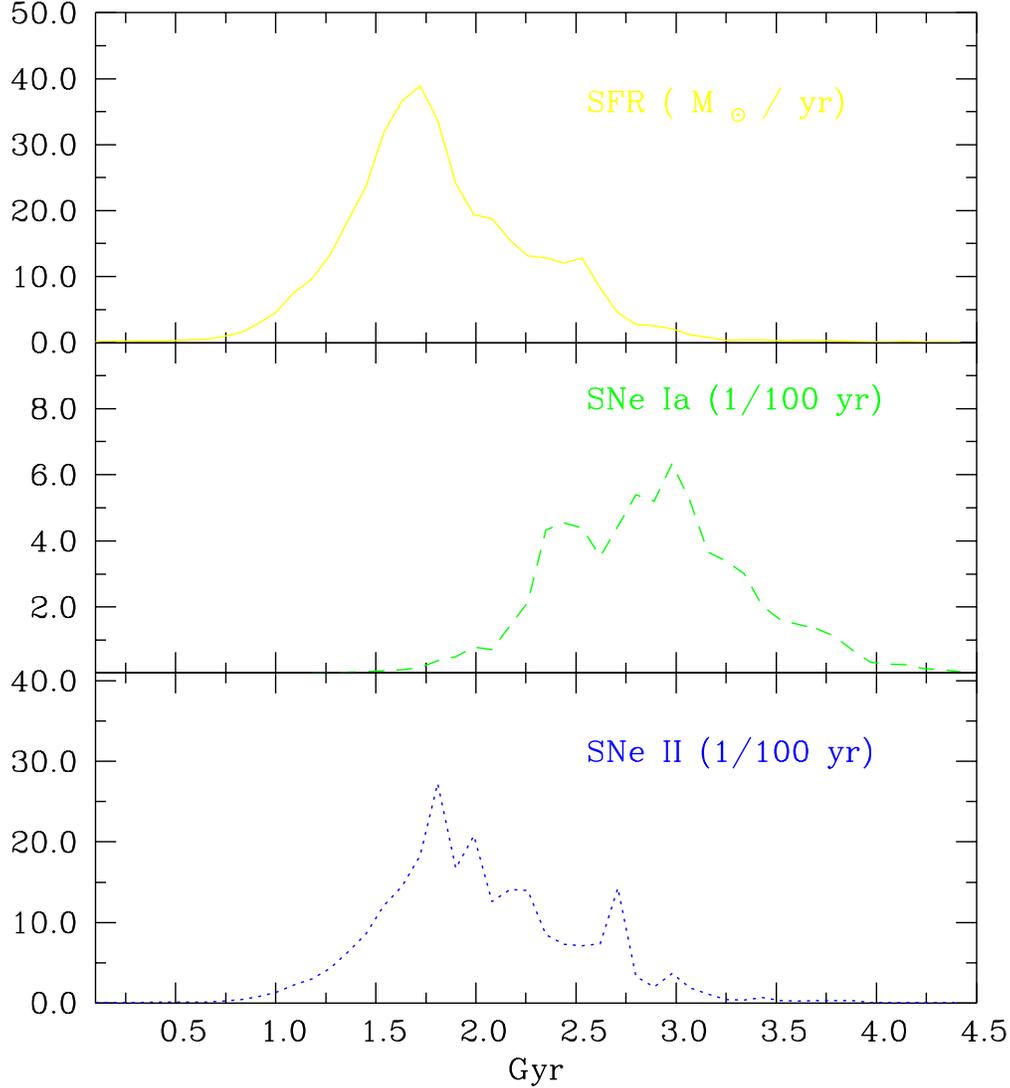,height=16cm,width=16cm}}
\caption{SF history and SN\ae\ rates for the model {\bf D3}. 
SN\ae\ II closely follow SF history, whereas SN\ae\  Ia exhibit
a peak somewhat later, as expected from the different progenitors
lifetime.}
\end{figure*}

\begin{figure*}
\centerline{\psfig{file=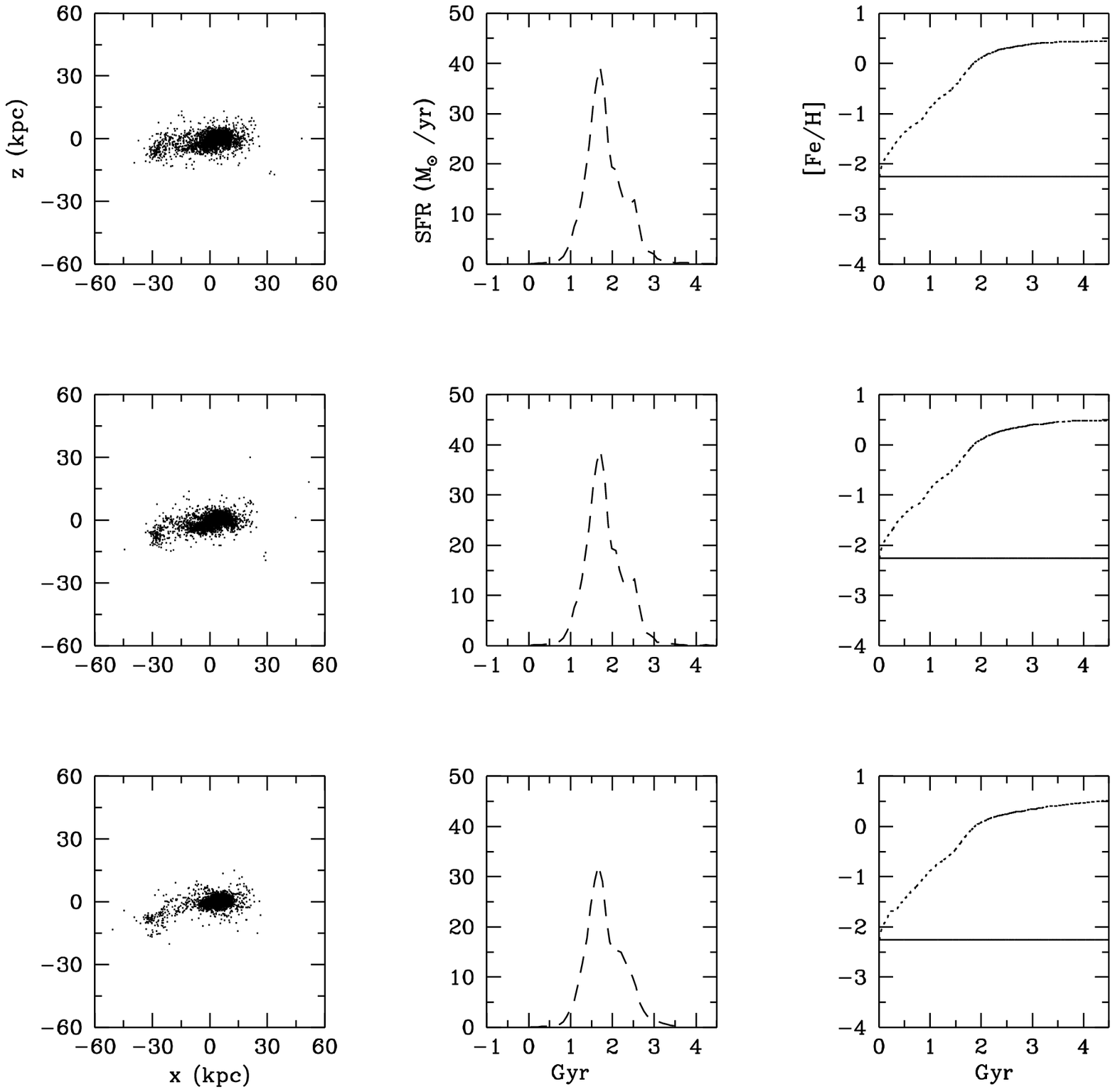,height=18cm,width=18cm}}
\caption{Comparison of the galactic models evolution by using different
sources of Feed-Back. The top row shows the evolution of a model in which
only SN\ae\ FB is considered, middle row a model in which also stellar winds are switched on,
and bottom row the case of a model in which all the FB sources are allowed to act.
Right panels show the chemical enrichment (solid line the maximum metallicity, dotted
line the minimum), middle panels the SF history, and
left panels the final morphology of the stellar disk.} 
\end{figure*}

\section{Stellar Feed-Back}
In this section, we test separately the effect of FB of different nature
making use of the case {\bf D}  for the recipe of star formation.
Furthermore, we include the effect of chemical enrichment. 

The energy FB  originates from  SN\ae\ explosions of both type Ia and II
(Greggio \& Renzini 1983), the  stellar winds from massive stars
(Chiosi \& Maeder 1986)
and the  UV flux from the same objects (Chiosi et al. 1998).

In order to evaluate the amount of energy injected into ISM by the above 
sources we suppose that  each newly formed star-particle, which for the mass 
resolution of our simulations has a total mass of the order of 
$10^7 M_{\odot}$, is actually made of a larger number of smaller 
subunits (the real stars) lumped together and distributed in mass according 
to a given initial mass function (IMF). For the purposes of the present
study we adopt the IMF by Miller \& Scalo 
(1979) over the mass interval from 0.1 to 120 $M_{\odot}$.

Chemical evolution is followed by means of  the closed-box approximation
(Tinsley 1980; Portinari et al 1998) and sharing of the metals among the gas-particles is 
described by means of a diffusive scheme (Groom 1997; Carraro et
al. 1998a).

Details on implementation of FB in our Tree-SPH code 
has been already reported 
in Carraro et al. (1998a),  to whom the reader should refer. 
With respect to the previous study, only two major changes have  been made. 
First, of the UV flux emitted by massive stars only a small fraction (0.01) is 
actually used. All the remaining UV flux is supposed to be re-processed by dust 
into the far-infrared and lost by the galaxy 
(Silva et al 1998). 
Second, the energy released to the ISM by a SN event  amounts only to
$\rm 10^{50} ergs$,
roughly 10 times less than in older evaluations (see Thornton et al. 1998
for an exhaustive discussion of this topic).
The heating rates  from the three sources of energy are displayed 
in Fig.~4 for the sake of illustration. A notable feature to remark, is that 
the energy injection by stellar winds may parallel that from SN\ae\ explosions, 
and the contribution from the UV flux plays a significant
role although a strong dumping factor has been applied. 
All the FB mechanisms have almost
the same trend, since they
derive basically from the same stellar sources.

\begin{table*}
\tabcolsep 0.50truecm 
\caption{Results for models with different Feed-Back.}
\begin{tabular}{lcccccc} \hline
\multicolumn{1}{c}{Case} &
\multicolumn{1}{c}{$T_{ave}$} &
\multicolumn{1}{c}{SF peak  } &
\multicolumn{1}{c}{Peak time} &
\multicolumn{1}{c}{$[Fe/H]_{fin}$} &
\multicolumn{1}{c}{$r_{D}$} & 
\multicolumn{1}{c}{$h_{D}$}\\
\hline
& $10^{6}~^{o}K$ & $M_{\odot} /yr$& $Gyr$& dex& kpc &kpc\\
\hline
{\bf D1}  & 1.15 & 39 & 1.72 & 0.446 & 30 & $5$\\
{\bf D2}  & 1.46 & 38 & 1.72 & 0.406 & 30 & $5$\\
{\bf D3}  & 1.33 & 32 & 1.67 & 0.513 & 20 & $5$\\
\hline
\end{tabular}
\end{table*}

Finally, we stress that all the energy from FB 
is given to the thermal budget of the particles, because the 
resolution we are working with makes it impossible to 
describe this process in more detail.
In fact, the space  resolution of these simulations
is about  2.5 kpc, much larger than
the typical distance over which the effects of
SN\ae\ explosions and stellar winds are visible
(Mckee \& Ostriker 1977). 
See also the arguments given by Carraro et al. (1998a). Possible 
interactions of kinetic nature among  gas particles occur 
only via the pressure gradients.

\subsection{Testing the sources of   Feed-Back}

In order to understand the role played by each source of FB
we perform the following experiments using the same model 
as in the previous section.  Three cases are considered:

\noindent
\begin{description}
\item{Case {\bf D1}: ~FB from SN\ae;}
\item{Case {\bf D2}: ~FB from SN\ae~ and stellar winds;}
\item{Case {\bf D3}: ~FB from SN\ae, stellar winds, and UV flux.}
\end{description}

\noindent
First of all in Fig.~5, for the case {\bf D1}, we show the SF history,
and the rate of SN\ae\ of both types, Ia and II.
Due to the different stellar progenitors, SN\ae\ Ia and II rates show  different
trends. SN\ae\ II explode immediately,
and their rate strictly follow the SF history. On the contrary, SN\ae\ Ia start to
appear later, being the progenitor less massive. The relative abundance of SN\ae\ events
clearly depends on the adopted IMF. 

The results of these simulations are shown in the various panels 
of  Fig.~6.  A few relevant quantities of the models are given in  Table~3.
Namely we consider the mean gas temperature, the SF peak and the time at which it occurs,
the final maximum metal abundance, 
and the radial and vertical dimensions of the final stellar disk.
 
Looking at these results the following conclusions can be drawn.
The models {\bf D1} and {\bf D2} do not differ significantly, simply 
turning on FB from stellar winds the final models becomes a bit
hotter, less stars are formed and the final maximum metal abundance
achieved is slightly lower. 
Since almost the same amount of stars are generated,
the final stellar disk morphology is basically identical.\\
When the combined effect of all the FB sources are considered
(model {\bf D3}) the situation changes more significantly
(see also Fig.~4).
The model becomes hotter more rapidly when SF starts,
SF peak occurs before, and less stars are formed. 
As a consequence the final mean temperature is somewhat lower 
and the disk morphology  is not well resolved as in the previous models.
The final maximum metallicity turns out somewhat greater due to the effect
of the diffusive mixing. Metals are spread and smooth around the gas 
neighbors. Since SF in this model stops before, no metal poor stars are 
produced, and the mean metallicity does increase.\\

In all the models a number of gas particles do not experience SF and
keep the primordial metal content.

\section {Summary and conclusions}

In the framework of the formation of a spiral-like 
galaxy, we have examined in detail the effect of different prescriptions
for Star Formation and Feed-Back on numerical
simulations of galaxy formation and evolution. \\

First of all we described in details how we integrate energy equation
when cooling is switched on. We adopted cooling 
functions which depend on metal abundance, 
and use the Brent method as root finder instead of the more widely used
Newton-Raphson scheme. Brent scheme performs better when using functions
in tabular form.\\

As far as the physical conditions at which Star Formation is likely to
start, we argue that the criterion on the  velocity divergence has
no sizable effect on the model results. \\
Moreover
the over-density and cooling time criteria are equivalent only if the effect of 
metallicity on the cooling time-scale is neglected. However, as in real 
galaxies the metallicity varies as a function of time and space, 
the criterion based
on the cooling time-scale ought to be preferred. This finding is a step further
toward better understanding under which conditions Star Formation takes place
in real galaxies.\\
In addition we introduce in the gas free-fall time computation the contribution of DM
and eventually stars which lie within the neighbors sphere. This has the effect to 
decrease the free-fall time, accelerating the cloud collapse.\\

Finally we have quantitatively shown the separated and cumulated effects of
different sources of Feed-Back (SN\ae\~, stellar wind and UV flux from massive stars) 
on the global properties  (SFR, 
metallicity and  morphology) of galaxy models.\\

\section*{Acknowledgements}
The authors deeply acknowledge an
anonymous referee for the detailed report on the first version of
the paper.
This study has been financed by several agencies: the Italian Ministry of
University, Scientific Research and Technology (MURST), the Italian
Space Agency (ASI), and the European Community (TMR grant ERBFMRX-CT-96-0086).

\end{document}